\documentclass[journal]{IEEEtran}
\usepackage{fancyhdr}
\usepackage{xcolor,soul,framed} 
\colorlet{shadecolor}{yellow}
\usepackage[pdftex]{graphicx}
\graphicspath{{../pdf/}{../jpeg/}}
\DeclareGraphicsExtensions{.pdf,.jpeg,.png}

\usepackage[cmex10]{amsmath}
\usepackage{array}
\usepackage{mdwmath}
\usepackage{mdwtab}
\usepackage{eqparbox}
\usepackage{url}
\usepackage{cite}
\usepackage{stfloats}
\renewcommand{\headrulewidth}{0pt}      
\renewcommand{\footrulewidth}{0pt}

\begin{document}

\bstctlcite{IEEEexample:BSTcontrol}

 \title{Electromagnetic Effective Degree of Freedom of a MIMO System in Free Space}
  \author{Shuai S. A. Yuan,~\IEEEmembership{Student Member,~IEEE,} Zi He,~\IEEEmembership{Senior Member,~IEEE,} Xiaoming Chen,~\IEEEmembership{Senior Member,~IEEE,} Chongwen Huang,~\IEEEmembership{Member,~IEEE,} and Wei E. I. Sha,~\IEEEmembership{Senior Member,~IEEE}
	\thanks{This research is funded by National Natural Science Foundation of China (Nos. 61975177, U20A20164). (Corresponding authors: Wei E. I. Sha and C. Huang)}
	\thanks{S. S. A. Yuan, C. Huang and W. E. I. Sha are with the College of Information Science and Electronic Engineering, Zhejiang University, Hangzhou 310027, China. (e-mails: weisha@zju.edu.cn and chongwenhuang@zju.edu.cn).}
	\thanks{Z. He is with the School of Electrical Engineering and Optical Technique, Nanjing University of Science and Technology, Nanjing 210094, China}
	\thanks{X. Chen is with the School of Information and Communications Engineering, Xi'an Jiaotong University, Xi'an 710049, China.}
}

\maketitle

\thispagestyle{fancy}         
\fancyhead{}       
\fancyfoot{}                
\lhead{Published in IEEE Antennas and Wireless Propagation Letters, https://doi.org/10.1109/LAWP.2021.3135018}        
\chead{}
\rhead{\thepage}
\lfoot{}
\rfoot{}
\renewcommand{\headrulewidth}{0pt}      
\renewcommand{\footrulewidth}{0pt}
\begin{abstract} %
Effective degree of freedom (EDOF) of a multiple-input-multiple-output (MIMO) system represents its equivalent number of independent single-input-single-output (SISO) systems, which directly characterizes the communication performance. Traditional EDOF only considers single polarization, where the full polarized components degrade into two independent transverse components under the far-field approximation. However, the traditional model is not applicable to complex scenarios especially for the near-field region. Based on an electromagnetic (EM) channel model built from the dyadic Green's function, we first calculate the EM EDOF to estimate the performance of an arbitrary MIMO system with full polarizations in free space. Then, we clarify the relations between the limit of EDOF and the optimal number of sources/receivers. Finally, potential benefits of near-field MIMO communications are demonstrated with the EM EDOF, in which the contribution of the longitudinally polarized source is taken into account. This work establishes a fundamental EM framework for MIMO wireless communications.
\end{abstract}


\IEEEpeerreviewmaketitle

\section{Introduction}

\IEEEPARstart{B}{ased} on the Shannon's information theory \cite{shannon}, multiple-input-multiple-output (MIMO) technology using spatial multiplexing has been developed for enhancing the channel capacity of modern wireless communications \cite{telatar1999capacity, TL2014}. Different electromagnetic (EM) modes, as orthogonal bases, have been employed in MIMO systems, like conventional plane-wave modes and newly-introduced orbital angular momentum modes \cite{yan2014high,kai2017orbital,zhang2016millimetre}. The sources/receivers in a MIMO system include antenna arrays \cite{TL2014} and metasurfaces \cite{David2019,cui2020}, represented by the discrete and quasi-continuous EM sources.

As an important concept in information theory, effective degree of freedom (EDOF) of a MIMO system represents its equivalent number of independent single-input-single-output (SISO) systems \cite{Kahn2000,muharemovic2008antenna}, which conveniently estimates the performance of the MIMO system through a single number. Compared to the concept of degree of freedom (DOF), which is defined as the number of significant singular values (eigenvalues) of the channel (correlation) matrix, the EDOF is directly related to the slope of spectral efficiency \cite{Verdu2002}, and thus to the channel performance. The EDOF of a MIMO system, as well as its limit, needs to be estimated with the channel matrix \cite{gesbert2002capacity, signal2005space}. Typically, the channel matrix is modeled based on the scalar Green's function for short-range MIMO in free space \cite{Ingram2005,Honma2009,Nam2011,hiraga2012effectiveness}, or some approximation methods tailored to multipath environments \cite{muharemovic2008antenna,RT2016}. However, these models either break down at the near-field region or cannot capture the full-wave physics (full polarizations) of EM fields.

As EM wave is the physical carrier of information, the DOF of MIMO system has also been investigated from EM perspectives\cite{Wallace2008,JR2006}. Mathematically, two groups of infinite orthogonal modes are defined in a Hilbert space to represent the source and radiation fields, then the DOF limit is deduced with functional methods \cite{MD2019,piestun2000electromagnetic}. To find the ultimate capacity of a MIMO antenna system \cite{Mats2021}, the surface currents of the antenna elements are expanded as the transmitting (source) bases. Furthermore, an interesting model for the hologram MIMO has been proposed recently \cite{Pizzo2020}, where the channel matrix is built with plane-wave expansion. Additionally, the DOF of a MIMO system is proved as the bound of its EDOF through EM methods \cite{Migliore2006}. Nevertheless, clear definition and detailed discussion of the EM EDOF of a MIMO system, as well as its connection with the EM DOF, have not been explored yet.

Three contributions are made in this work. Firstly, with the aid of a mathematical definition of EDOF from information theory, the EM EDOF is defined based on an EM channel matrix deduced from the dyadic Green's function. Secondly, the DOF and EDOF limit of a space-constrained MIMO system are discussed. Thirdly, we demonstrate the potential benefits of near-field MIMO communications with the consideration of the longitudinally polarized current source. These comprehensive and comparative studies have not been investigated before.
%
\section{EM channel model of a MIMO system}
\begin{figure}[ht!]
	\centering
	\includegraphics[width=3.4in]{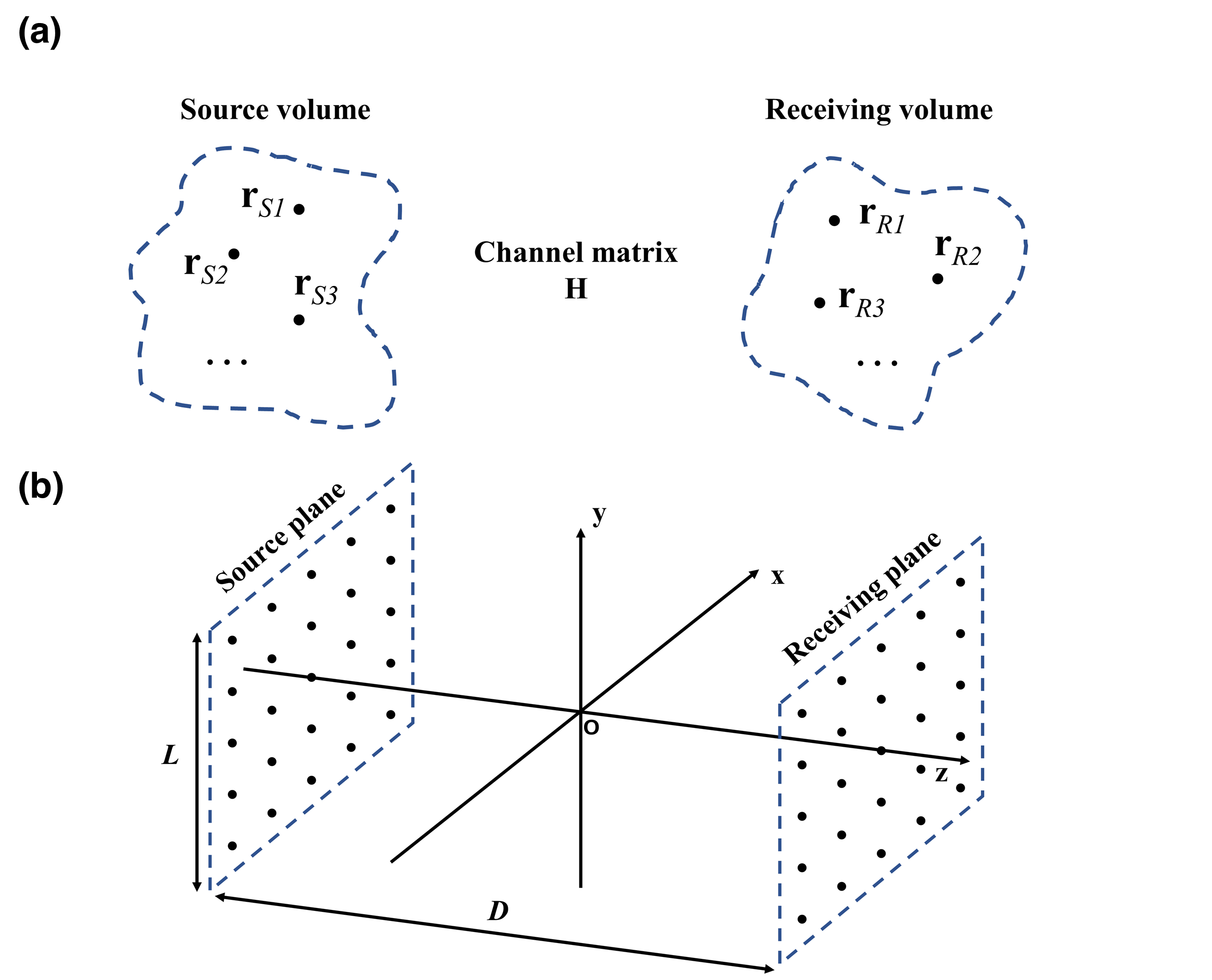}
	\caption{(a) A basic model of a MIMO system in free space. Transmitting antennas at the positions $\mathbf{r}_{S n}\left(n=1, \ldots, N_{S}\right)$ are distributed in the source volume and receiving antennas at the positions $\mathbf{r}_{R m}\left(m=1, \ldots, N_{R}\right)$ are distributed in the receiving volume. The transmitting/receiving antennas in the two volumes are coupled by a channel matrix $\mathbf{H}$. (b) A MIMO system consists of $N\times N$ uniformly distributed point sources/receivers on two identical square source/receiving planes. $L$ is the side length of the planes, and $D$ is the distance between the two planes.}
	\label{MIMO system}
\end{figure}
Considering a basic MIMO model in free space, as depicted in Fig. 1 (a), a set of $N_S$ transmitting antennas at the positions $\mathbf{r}_{S n}\left(n=1, \ldots, N_{S}\right)$ are distributed in the source volume, $N_R$ receiving antennas at the positions $\mathbf{r}_{R m}\left(m=1, \ldots, N_{R}\right)$ are distributed in the receiving volume. To make the following equations clear, the transmitting and receiving antennas are modeled as isotropic point sources/receivers (delta function basis), which is the widely-used assumption made in EM information theory \cite{miller2019waves}. From EM perspective, the antennas can also be modeled as continuous surface (equivalent) currents by rooftop or Rao-Wilton-Glisson (RWG) basis, as frequently utilized in the methods of moments \cite{gibson2007method}. Different basis representations of the currents, which relates to different antenna designs, will not influence the estimations of the EDOF and DOF limit. Therefore, the mutual coupling between antenna elements is ignored here to draw fundamental physical conclusions and explore insightful engineering rules.

\subsection{Single-polarization model}
For a set of $N_S$ point sources at the positions $\mathbf{r}_{S n}$ in the source volume with the complex amplitudes $t_n$, as shown in Fig. 1 (a), the superposed electric field at the positions $\mathbf{r}_{Rm}$ in the receiving volume would be
\begin{equation}
E\left(\mathbf{r}_{R m}\right)=\frac{1}{4 \pi} \sum_{n=1}^{N_{S}} \frac{\exp \left(-j k_0\left|\mathbf{r}_{R m}-\mathbf{r}_{S n}\right|\right)}{\left|\mathbf{r}_{R m}-\mathbf{r}_{S n}\right|} t_{n}=\sum_{n=1}^{N_{S}} h_{mn} t_{n},
\end{equation}
where
\begin{equation}
h_{mn}=\frac{1}{4 \pi} \frac{\exp \left(-j k_0\left|\mathbf{r}_{R m}-\mathbf{r}_{S n}\right|\right)}{\left|\mathbf{r}_{R m}-\mathbf{r}_{S n}\right|}=g\left(\mathbf{r}_{R m}, \mathbf{r}_{Sn}\right),
\end{equation}
which is the scalar Green's function with certain source/receiver positions. The received signals at the point receivers would be the sum of the fields from all the point sources added up at $\mathbf{r}_{Rm}$
\begin{equation}
f_{m}=\sum_{n=1}^{N_{S}} h_{mn} t_{n}.
\end{equation}
If $ t_{n}$ and $f_{m}$ are collected in the two column vectors ${t}=[t_1, t_2, \dots, t_{N_S}]^T$ and ${f}=[f_1, f_2, \dots, f_{N_R}]^T$, we can define the projection from the point sources to the point receivers as
\begin{equation}
{f}=\mathbf{H} {t},
\end{equation}
 with
\begin{equation}
\mathbf{H}=\left[\begin{array}{cccc}
h_{11} & h_{12} & \cdots & h_{1 N_{S}} \\
h_{21} & h_{22} & \cdots & h_{2 N_{S}} \\
\vdots & \vdots & \ddots & \vdots \\
h_{N_{R} 1} & h_{N_{R} 2} & \cdots & h_{N_{R} N_{S}}
\end{array}\right],
\end{equation}
which is the channel matrix based on the scalar wave equation. This model considering single polarization is sufficient for far-field communications, but not for near-field communications.

\subsection{Full-polarization model}
In order to model an arbitrary MIMO system with full polarizations, dyadic Green's function is utilized. Considering the same configuration as the previous subsection, we adopt three polarizations at each point source/receiver. The dyadic Green's function in free space is
\begin{equation}
\bar{{\mathbf{G}}}\left(\mathbf{r}, \mathbf{r}^{\prime}\right)=\left(\bar{{\mathbf{I}}}+\frac{\nabla \nabla}{k_{0}^{2}}\right) g\left(\mathbf{r}, \mathbf{r}^{\prime}\right),
\end{equation}
where $\bar{{\mathbf{I}}}$ is the unit tensor. The tensor $\bar{{\mathbf{G}}}$ could be rewritten as a matrix form, i.e.
\begin{equation}
\bar{{\mathbf{G}}}=\left[\begin{array}{ccc}
G_{xx} & G_{xy}  & G_{xz} \\
G_{yx} & G_{yy}  & G_{yz} \\
G_{zx} & G_{zy}  & G_{zz}
\end{array}\right],
\end{equation}
\noindent where each element is a scalar Green's function between one polarization of field and one polarization of source, denoted by its subscript.

As discussed in the previous subsection, an $N_R$ $\times$ $N_S$ channel matrix is built for a MIMO system with one scalar Green's function. The three polarizations of electric field are orthogonal, so that we could write the channel of full polarizations in a matrix form without loss of information. Considering Eq. (4) in the full-polarization case, the complex amplitudes of sources ${t}=[{t}_x {t}_y {t}_z]^T=[t_{x1}, \cdots, t_{xN_S}, t_{y1}, \cdots, t_{yN_S}, t_{z1}, \cdots, t_{zN_S}]^T$ is a 3$N_S$ $\times$ 1 vector,  the complex amplitudes of received signals ${f}=[{f}_x {f}_y {f}_z]^T=[f_{x1}, \cdots, f_{xN_R}, f_{y1}, \cdots, f_{yN_R}f_{z1}, \cdots, f_{zN_R}]^T$ is a 3$N_R$ $\times$ 1 vector, where the $x, y, z$ in subscripts denote the polarizations and the $1, 2, \cdots, N$ in subscripts denote the positions. The two column vectors are then related by a 3$N_R$ $\times$ 3$N_S$ EM channel matrix
\begin{equation}
\mathbf{H}=\left[\begin{array}{ccc}
\mathbf{H}_{xx} & \mathbf{H}_{xy}  & \mathbf{H}_{xz} \\
\mathbf{H}_{yx} & \mathbf{H}_{yy}  & \mathbf{H}_{yz} \\
\mathbf{H}_{zx} & \mathbf{H}_{zy}  & \mathbf{H}_{zz}
\end{array}\right],
\end{equation}
\begin{figure}[ht!]
	\centering
	\includegraphics[width=2.5in]{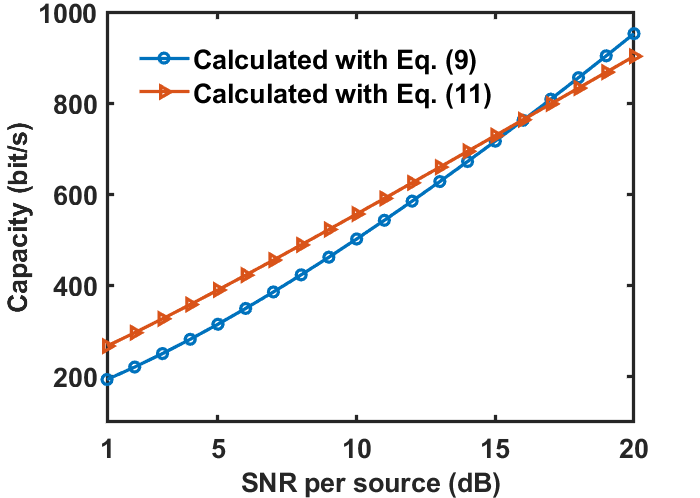}
	\caption{Comparison between the capacities calculated with Eq. (9) and Eq. (11). 20 $\times$ 20 point sources/receivers are uniformly distributed on 10 $\times$ 10 $\lambda_0^2$ source/receiving plane, the distance between the two planes is 7 $\lambda_0$, $\lambda_0$ is free-space wavelength, and bandwidth $B$ is set as 1.}
	\label{MIMO}
\end{figure}
\par\noindent where the nine $N_R$ $\times$ $N_S$ matrices correspond to the nine scalar Green's functions. This matrix contains all the information needed for a MIMO system in free space, and the model could be extended to arbitrary expansion bases and propagation environments. Notice that the correlation matrix $ \mathcal{R} ={\mathbf{H}}{\mathbf{H}}^\dagger$ or $ {{\mathbf{H}}^\dagger\mathbf{H}}$ sharing the same eigenvalues is used for estimating the performance of a MIMO system, as the receiving (transmitting) power is related by $\bar{\mathbf{G}}^\dagger\bar{\mathbf{G}}$ ($\bar{\mathbf{G}}\bar{\mathbf{G}}^\dagger$) \cite{piestun2000electromagnetic}.

\section{EM EDOF of a MIMO system}
\subsection{Definition of EM EDOF}
The concept of the EDOF of a MIMO system originates from information theory \cite{muharemovic2008antenna,Kahn2000,Verdu2002}, which indicates the spectral efficiency (channel capacity at a single frequency) of a MIMO system through a scalar number. Traditional EDOF is calculated with Eq. (5), thus cannot be applicable to the near-field regime. The EM EDOF introduced here utilizes the same mathematical operation, but the EM channel matrix, i.e., Eq. (8), incorporates full polarizations.

Considering that the power is equally allocated to identical antenna elements with fixed total signal-to-noise ratio (SNR) $\rho$, which is commonly adopted in typical MIMO applications. The capacity of a MIMO system can be calculated with
\begin{equation}
C=B \sum_{i=1}^{n} \log _{2}\left(1+\frac{\rho}{n} \sigma_{i}\right),
\end{equation}
where $B$ is the antenna bandwidth, $n$ is the rank (or DOF) of the correlation matrix $\mathcal{R}$, and $\sigma_{i}$ are the corresponding eigenvalues. If the MIMO system is ideal ($\mathcal{R}$ is a unit matrix, all the subchannels are independent), Eq. (8) becomes
\begin{equation}
C=B n \log _{2}\left(1+\frac{\rho}{n} \right).
\end{equation}
This ideal case is unreachable because $\sigma_{i}$ are different in practical. However, in reality, the channel capacity can still be written as a separated form
\begin{equation}
C\approx B \Psi_e \log _{2}\left(1+\frac{\rho}{\Psi_e} \right),
\end{equation}
where $\Psi_e \in$ [1, n]  is the EDOF, i.e., the equivalent number of SISO systems, and is equal to $n$ in ideal case. The concept and estimation of the EDOF have been discussed in \cite{muharemovic2008antenna,Kahn2000,Verdu2002}, and $\Psi_e$ can be approximately calculated as
\begin{figure}[ht!]
	\centering
	\includegraphics[width=3.4in]{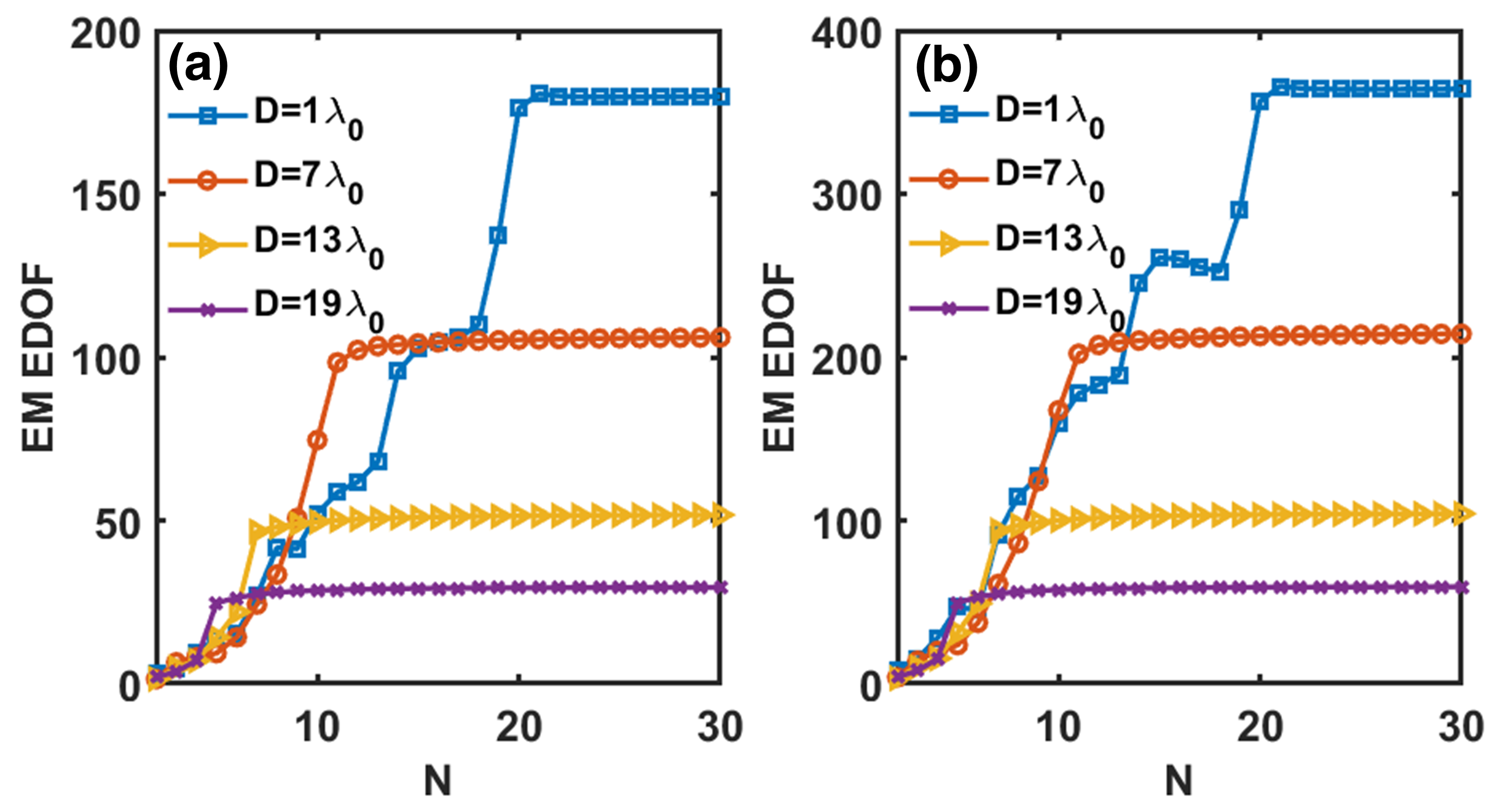}
	\caption{EM EDOFs calculated with different $N$ (number of sources/receivers in one side) and $D$ (distance). Side length $L$ is fixed as 10$\lambda_0$. {(a) Scalar Green's function. (b) Dyadic Green's function.} }
	\label{Near far field}
\end{figure}
\begin{equation}\label{DoF}
\Psi_e\left(\mathcal{R}\right)=\left(\frac{\operatorname{tr}\left(\mathcal{R}\right)}{\left\|\mathcal{R}\right\|_{F}}\right)^{2}=\frac{\left(\sum_{i} \sigma_{i}\right)^{2}}{\sum_{i} \sigma_{i}^{2}},
\end{equation}
\noindent where $\operatorname{tr}(\cdot)$ represents the trace operator and the subscript $F$ denotes the Frobenius norm, this equation has been proved to be sufficiently accurate in estimating the capacity of a practical MIMO system, as seen in Fig. 2.
\subsection{Limit of EM EDOF}
 In practical implementation, engineers are concerned about the limit of EM EDOF of a space-constrained MIMO system, and also the number of sources/receivers for approaching the limit. Intuitively, when the number of sources/receivers surpasses a certain saturation point, the exceeding eigenvalues, corresponding to the exceeding sources/receivers, will become too small to contribute to the EM EDOF, and the EM EDOF will reach its maximum value. With the proposed EM channel model, we can precisely estimate the limit of EM EDOF of a MIMO system.

Without loss of generality, we consider a MIMO system consisting of two $L\times L$ square planes separated by the distance $D$, with uniformly distributed $N\times N$ point sources/receivers on the source/receiver planes, as depicted in Fig. 1 (b). To investigate the limit of EM EDOF, $L$ is fixed as 10$\lambda_0$, then the EM EDOFs with different $N$ and distance $D$ are calculated. The results calculated with scalar and dyadic Green's functions are presented in Fig 3. Regarding the scalar Green's function case, the far-field E-field is polarized at $\emph{x}$ or $\emph{y}$ direction shown in Fig. 1 (b). It can be observed that when $N$ increases, the EM EDOF will increase fast in the beginning, then reach a maximum value in all the cases. An interesting phenomenon is that the EDOF appears a partial decrease at $D = 1\lambda_0$ in Fig. 3 (b), while similar tendency does not appear in the scalar case. The reason is that the dyadic Green's function produces a rigorous wave-physics solution to capture the near-field evanescent waves with three polarized components.%

\subsection{Optimal number of sources/receivers}
Optimal number of sources/receivers is simultaneously estimated in the previous subsection, which is the number of significant eigenvalues (i.e., DOF) of the EM correlation matrix. Nonetheless, there are some elegant intuitive methods. 
\begin{figure}[ht!]
	\centering
	\includegraphics[width=2.5in]{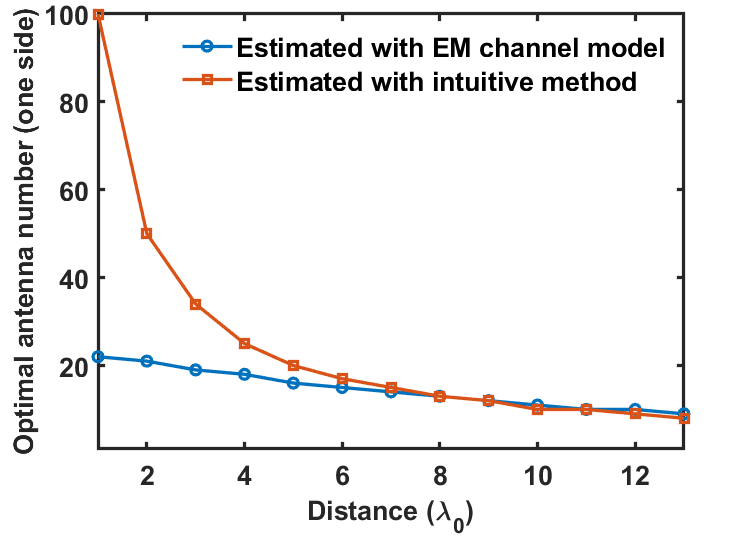}
	\caption{Optimal number of sources/receivers calculated with the EM channel model and the intuitive method, with the side length $L= 10\lambda_0$ and distance $D = 1-13\lambda_0$.}
	\label{Near far field}
\end{figure}
Under the paraxial approximation in optics, i.e., the distance between the two planes is much larger than their sizes, an intuitive method from the perspective of solid angle is proposed to calculate the number of significant EM modes\cite{miller2000communicating}

\begin{equation}
\Psi=\frac{A_{S} A_{R}}{\lambda_0^{2} D^{2}},
\end{equation}
where $A_S$ is the area of source plane, $A_R$ is the area of receiving plane and $D$ is the distance between the two planes. We compare the optimal number of sources/receivers calculated with the EM channel model and the intuitive method, with $L= 5\lambda_0$ and $D = 1-13\lambda_0$. As shown in Fig. 4, when the distance between the two planes becomes large, the two methods give almost the same results, indicating that the intuitive method is accurate at a relatively long distance. Eq. (13) can also be extended to half-space isotropic multipath environment with a source plane and an ideal half-space receiving plane, also regarded as a Fourier spectral method \cite{Pizzo2020, Davide2020,loyka2004information}. The intuitive methods are only suitable for the MIMO systems at far field, while the EM channel model could be readily applicable to more complex scenarios. 

The EM DOF of a MIMO system refers to the number of dominant EM modes (rank), thus it is identical to the optimal number of sources/receivers for approaching the limit of EM EDOF. The EM EDOF is rather a mathematical tool to denote the equivalent number of SISOs of a MIMO system, which is directly related to the spectral efficiency.
\section{Near-field MIMO communications}
It is interesting to investigate the potential benefit of the near-field (reactive) MIMO communications compared to the far-field communications with the EM EDOF, which cannot be accomplished with the traditional method. Other than the enhancement of EM EDOF brought by the decrease of distance, the available polarizations of current sources are fundamentally different at near field and far field. Considering the dyadic Green's function with the far-field approximation $\nabla \rightarrow j \mathbf{k}_0$, Eq. (6) becomes
\begin{equation}
\bar{{\mathbf{G}}}^f\left(\mathbf{r}, \mathbf{r}^{\prime}\right)=\left(\bar{{\mathbf{I}}}-\mathbf{a_r}\mathbf{a_r}\right) g\left(\mathbf{r}, \mathbf{r}^{\prime}\right),
\end{equation}
\begin{figure}[ht!]
	\centering
	\includegraphics[width=2.5in]{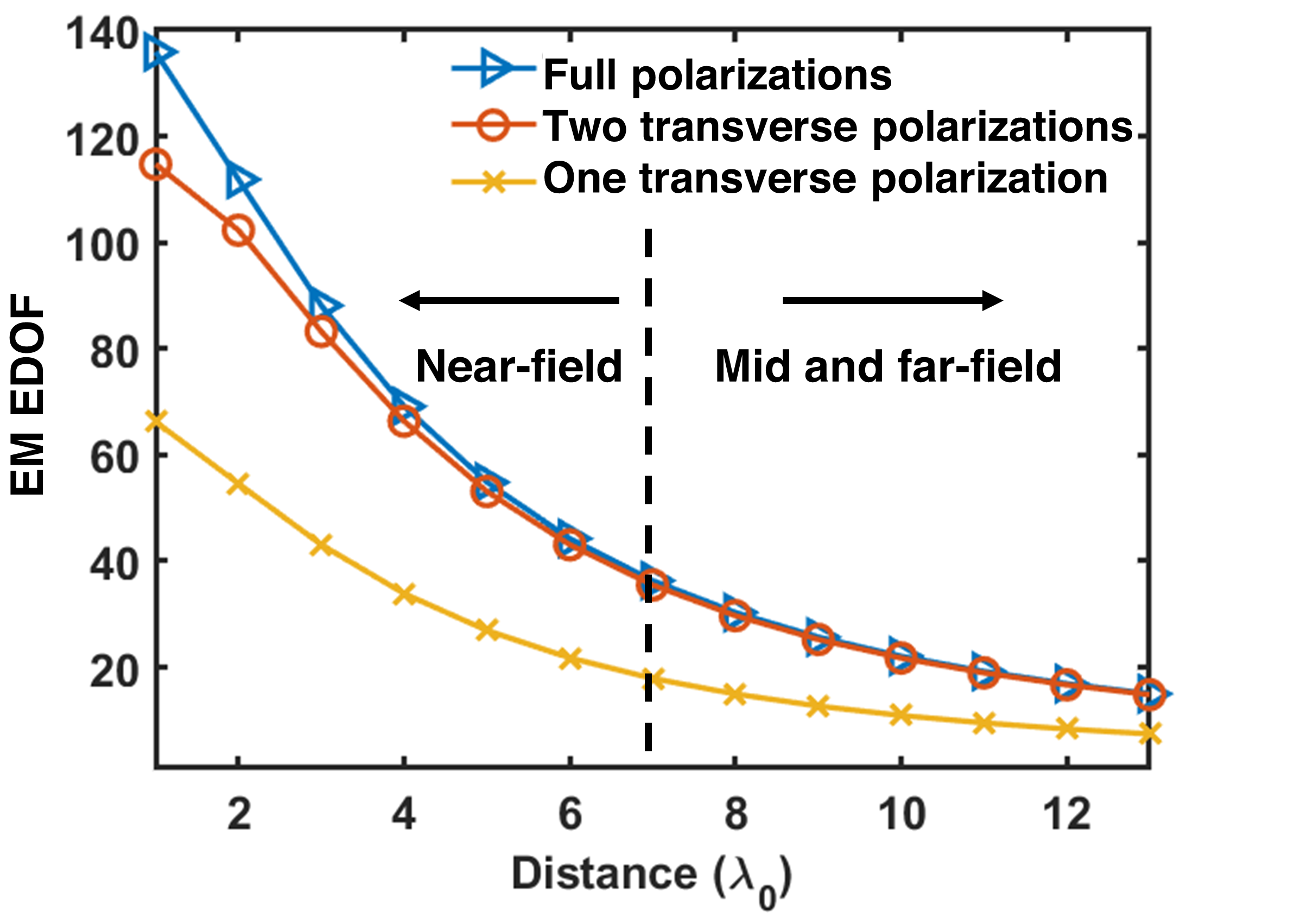}
	\caption{EM EDOFs of MIMO systems calculated with different number of polarizations, the side length $L= 5\lambda_0$, antenna number $N=11$ and distance $D = 1-13\lambda_0$.}
	\label{Near far field}
\end{figure}
where $\mathbf{a_r}\mathbf{a_r}$ is the $\emph{r}$-polarized component of unit tensor in spherical coordinate, i.e., the $\emph{z}$-polarized component in the Cartesian coordinate in Fig. 1 (b). Obviously, the $\emph{z}$ polarization is subtracted, which indicates that the $\emph{z}$ polarization of current source will not contribute to the far-field communications. Therefore, only the two transverse polarizations of current sources are retained, the dyadic Green's function with the far-field approximation can be simplified to be
\begin{equation}
\bar{{\mathbf{G}}}^f=\left[\begin{array}{ccc}
G_{xx} & G_{xy}  & 0 \\
G_{yx} & G_{yy}  & 0 \\
G_{zx} & G_{zy}  & 0
\end{array}\right],
\end{equation}
where the third column denotes the contribution of the $\emph{z}$ polarization.

In order to demonstrate the potential benefit of near-field MIMO communications, we calculate the EM EDOF based on the dyadic Green's functions with one transverse, two transverse and full polarizations, i.e., Eq. (15) and Eq. (7) for the latter two, respectively. With the side length $L= 5\lambda_0$, the regions of near-field and far-field can be calculated as $D_n = 0.62\cdot(L^3/\lambda_0)^{1/2} = 6.9\lambda_0$ and $D_f = 2\cdot L^2/\lambda_0= 50\lambda_0$ \cite{balanis2015antenna}. The number of antennas is fixed as 11$\times$11, with which the limit of EM EDOF is almost approached in all the cases. The contribution of the $\emph{z}$ polarization of current source at near field is clearly illustrated in Fig. 5. It can be observed that the benefit brought by the $\emph{z}$ polarization vanishes when the distance $D$ exceeds the near-field range; the EM EDOF of full polarizations doubles that of single polarization at mid and far fields. Hence, the benefit of near-field MIMO communications from the $\emph{z}$ polarization only appears at a relatively small distance in free space.

\section{Conclusion}
Based on the EM channel model of a MIMO system in free space, we discussed the principle of EM EDOF, limit of EM EDOF and the optimal number of sources/receivers. This method is useful for estimating the capacity limit of a MIMO system from EM perspective, and can also be utilized for optimizing array arrangement as well as analyzing near-field communication. The proposed method can be readily extended to arbitrary source/receiver types and propagating environments, by using proper basis functions \cite{gibson2007method} and numerical dyadic Green's function in arbitrary background \cite{Chew2020}.

\ifCLASSOPTIONcaptionsoff
  \newpage
\fi
\bibliographystyle{IEEEtran}
\bibliography{IEEEabrv,Bibliography}

\begin{thebibliography}{10}
\providecommand{\url}[1]{#1}
\csname url@rmstyle\endcsname
\providecommand{\newblock}{\relax}
\providecommand{\bibinfo}[2]{#2}
\providecommand\BIBentrySTDinterwordspacing{\spaceskip=0pt\relax}
\providecommand\BIBentryALTinterwordstretchfactor{4}
\providecommand\BIBentryALTinterwordspacing{\spaceskip=\fontdimen2\font plus
\BIBentryALTinterwordstretchfactor\fontdimen3\font minus
  \fontdimen4\font\relax}
\providecommand\BIBforeignlanguage[2]{{%
\expandafter\ifx\csname l@#1\endcsname\relax
\typeout{** WARNING: IEEEtran.bst: No hyphenation pattern has been}%
\typeout{** loaded for the language `#1'. Using the pattern for}%
\typeout{** the default language instead.}%
\else
\language=\csname l@#1\endcsname
\fi
#2}}
\renewcommand\BIBentryALTinterwordstretchfactor{4}

\bibitem{shannon}
C.~E. Shannon, ``A mathematical theory of communication,'' \emph{Bell system
  technical journal}, vol.~27, no.~3, pp. 379--423, 1948.

\bibitem{telatar1999capacity}
E.~Telatar, ``Capacity of multi-antenna gaussian channels,'' \emph{Eur. Trans.
  Telecomm.}, vol.~10, no.~6, pp. 585--595, 1999.

\bibitem{TL2014}
E.~G. Larsson, O.~Edfors, F.~Tufvesson, and T.~L. Marzetta, ``Massive {MIMO}
  for next generation wireless systems,'' \emph{IEEE Commun. Mag.}, vol.~52,
  no.~2, pp. 186--195, 2014.

\bibitem{yan2014high}
Y.~Yan, G.~Xie, M.~P. Lavery, H.~Huang, N.~Ahmed, C.~Bao, Y.~Ren, Y.~Cao,
  L.~Li, Z.~Zhao, \emph{et~al.}, ``High-capacity millimetre-wave communications
  with orbital angular momentum multiplexing,'' \emph{Nat. Commun.}, vol.~5,
  no.~1, pp. 1--9, 2014.

\bibitem{kai2017orbital}
C.~Kai, P.~Huang, F.~Shen, H.~Zhou, and Z.~Guo, ``Orbital angular momentum
  shift keying based optical communication system,'' \emph{IEEE Photonics J.},
  vol.~9, no.~2, pp. 1--10, 2017.

\bibitem{zhang2016millimetre}
C.~Zhang and L.~Ma, ``Millimetre wave with rotational orbital angular
  momentum,'' \emph{Sci Rep}, vol.~6, no.~1, pp. 1--8, 2016.

\bibitem{David2019}
N.~Shlezinger, O.~Dicker, Y.~C. Eldar, I.~Yoo, M.~F. Imani, and D.~R. Smith,
  ``Dynamic metasurface antennas for uplink massive mimo systems,'' \emph{IEEE
  Trans. Commun.}, vol.~67, no.~10, pp. 6829--6843, 2019.

\bibitem{cui2020}
W.~Tang, J.~Y. Dai, M.~Z. Chen, K.-K. Wong, X.~Li, X.~Zhao, S.~Jin, Q.~Cheng,
  and T.~J. Cui, ``Mimo transmission through reconfigurable intelligent
  surface: System design, analysis, and implementation,'' \emph{IEEE J. Sel.
  Areas Commun.}, vol.~38, no.~11, pp. 2683--2699, 2020.

\bibitem{Kahn2000}
D.-S. Shiu, G.~Foschini, M.~Gans, and J.~Kahn, ``Fading correlation and its
  effect on the capacity of multielement antenna systems,'' \emph{IEEE Trans.
  Commun.}, vol.~48, no.~3, pp. 502--513, 2000.

\bibitem{muharemovic2008antenna}
T.~Muharemovic, A.~Sabharwal, and B.~Aazhang, ``Antenna packing in low-power
  systems: Communication limits and array design,'' \emph{IEEE Trans. Inf.
  Theory}, vol.~54, no.~1, pp. 429--440, 2008.

\bibitem{Verdu2002}
S.~Verdu, ``Spectral efficiency in the wideband regime,'' \emph{IEEE Trans.
  Inf. Theory}, vol.~48, no.~6, pp. 1319--1343, 2002.

\bibitem{gesbert2002capacity}
D.~Gesbert, T.~Ekman, and N.~Christophersen, ``Capacity limits of dense
  palm-sized {MIMO} arrays,'' in \emph{Global Telecommunications Conference,
  2002. GLOBECOM'02. IEEE}, vol.~2, 2002, pp. 1187--1191.

\bibitem{signal2005space}
A.~Poon, R.~Brodersen, and D.~Tse, ``Degrees of freedom in multiple-antenna
  channels: a signal space approach,'' \emph{IEEE Trans. Inf. Theory}, vol.~51,
  no.~2, pp. 523--536, 2005.

\bibitem{Ingram2005}
J.-S. Jiang and M.~Ingram, ``Spherical-wave model for short-range mimo,''
  \emph{IEEE Trans. Commun.}, vol.~53, no.~9, pp. 1534--1541, 2005.

\bibitem{Honma2009}
N.~Honma, K.~Nishimori, T.~Seki, and M.~Mizoguchi, ``Short range mimo
  communication,'' in \emph{2009 3rd European Conference on Antennas and
  Propagation}, 2009, pp. 1763--1767.

\bibitem{Nam2011}
Y.~Tak and S.~Nam, ``Mode-based computation method of channel characteristics
  for a near-field mimo,'' \emph{IEEE Antennas Wirel. Propag. Lett.}, vol.~10,
  pp. 1170--1173, 2011.

\bibitem{hiraga2012effectiveness}
K.~Hiraga, T.~Seki, K.~Nishimori, and K.~Uehara, ``Effectiveness of short-range
  mimo using dual-polarized antenna,'' \emph{IEICE Trans. Commun.}, vol.~95,
  no.~1, pp. 87--96, 2012.

\bibitem{RT2016}
S.~Biswas, C.~Masouros, and T.~Ratnarajah, ``Performance analysis of large
  multiuser {MIMO} systems with space-constrained 2-d antenna arrays,''
  \emph{IEEE Trans. Wirel. Commun.}, vol.~15, no.~5, pp. 3492--3505, 2016.

\bibitem{Wallace2008}
M.~A. Jensen and J.~W. Wallace, ``Capacity of the continuous-space
  electromagnetic channel,'' \emph{IEEE Trans. Antennas Propag.}, vol.~56,
  no.~2, pp. 524--531, 2008.

\bibitem{JR2006}
J.~Xu and R.~Janaswamy, ``Electromagnetic degrees of freedom in 2-d scattering
  environments,'' \emph{IEEE Trans. Antennas Propag.}, vol.~54, no.~12, pp.
  3882--3894, 2006.

\bibitem{MD2019}
M.~D. Migliore, ``Horse (electromagnetics) is more important than horseman
  (information) for wireless transmission,'' \emph{IEEE Trans. Antennas
  Propag.}, vol.~67, no.~4, pp. 2046--2055, 2019.

\bibitem{piestun2000electromagnetic}
R.~Piestun and D.~A. Miller, ``Electromagnetic degrees of freedom of an optical
  system,'' \emph{J. Opt. Soc. Am. A-Opt. Image Sci. Vis.}, vol.~17, no.~5, pp.
  892--902, 2000.

\bibitem{Mats2021}
C.~Ehrenborg, M.~Gustafsson, and M.~Capek, ``Capacity bounds and degrees of
  freedom for {MIMO} antennas constrained by q-factor,'' \emph{IEEE Trans.
  Antennas Propag.}, pp. 1--1, 2021.

\bibitem{Pizzo2020}
A.~Pizzo, T.~L. Marzetta, and L.~Sanguinetti, ``Spatially-stationary model for
  holographic {MIMO} small-scale fading,'' \emph{IEEE J. Sel. Areas Commun.},
  vol.~38, no.~9, pp. 1964--1979, 2020.

\bibitem{Migliore2006}
M.~Migliore, ``On the role of the number of degrees of freedom of the field in
  mimo channels,'' \emph{IEEE Trans. Antennas Propag.}, vol.~54, no.~2, pp.
  620--628, 2006.

\bibitem{miller2019waves}
D.~A. Miller, ``Waves, modes, communications, and optics: a tutorial,''
  \emph{Adv. Opt. Photonics}, vol.~11, no.~3, pp. 679--825, 2019.

\bibitem{gibson2007method}
W.~C. Gibson, \emph{The method of moments in electromagnetics}, 2007.

\bibitem{miller2000communicating}
D.~A. Miller, ``Communicating with waves between volumes: evaluating orthogonal
  spatial channels and limits on coupling strengths,'' \emph{Appl. Optics},
  vol.~39, no.~11, pp. 1681--1699, 2000.

\bibitem{Davide2020}
D.~Dardari, ``Communicating with large intelligent surfaces: Fundamental limits
  and models,'' \emph{IEEE J. Sel. Areas Commun.}, vol.~38, no.~11, pp.
  2526--2537, 2020.

\bibitem{loyka2004information}
S.~Loyka, ``Information theory and electromagnetism: Are they related?'' in
  \emph{2004 10th International Symposium on Antenna Technology and Applied
  Electromagnetics and URSI Conference}, 2004, pp. 1--5.

\bibitem{balanis2015antenna}
C.~A. Balanis, \emph{Antenna theory: analysis and design}, 2015.

\bibitem{Chew2020}
H.~H. Gan, Q.~I. Dai, T.~Xia, W.~C. Chew, and C.-F. Wang, ``Broadband spectral
  numerical green's function for electromagnetic analysis of inhomogeneous
  objects,'' \emph{IEEE Antennas Wirel. Propag. Lett.}, vol.~19, no.~7, pp.
  1063--1067, 2020.

\end{thebibliography}

\vfill


\end{document}